\documentclass[amsmath,amssymb, aps, prl, twocolumn, notitlepage]{revtex4-2}
\usepackage{graphicx}% Include figure files
\usepackage{dcolumn}% Align table columns on decimal point
\usepackage{bm}
\usepackage[usenames]{color}
\usepackage{slashed}
\usepackage[utf8]{inputenc}
\usepackage{amsfonts}
\usepackage{amsmath}
\usepackage{amssymb}
\usepackage{hyperref}
\usepackage{latexsym}
\usepackage{color}
\usepackage{mhchem}
\usepackage{braket}
\usepackage{diagbox}
\usepackage{booktabs}
\usepackage{MnSymbol}
\usepackage{float}
\makeatletter
\let\newfloat\newfloat@ltx
\usepackage{algorithm}
\usepackage{algorithmicx}
\usepackage{algpseudocode}
\usepackage{multirow}
\newcommand{\printfnsymbol}[1]{%
  \textsuperscript{\@fnsymbol{#1}}%
}

\begin{document}
\title{The impact of dimensionality on universality of quantum Hall transitions}

\author{Qiwei Wan}
\affiliation{International Center for Quantum Materials, School of Physics, Peking University, Beijing, 100871, China}

\author{Yi Zhang}
\email{frankzhangyi@pku.edu.cn}
\affiliation{International Center for Quantum Materials, School of Physics, Peking University, Beijing, 100871, China}

\date{\today}

\begin{abstract}
Regardless of model and platform details, the critical phenomena exhibit universal behaviors that are remarkably consistent across various experiments and theories, resulting in a significant scientific success of condensed matter physics. One widely known and commonly used example is the 2D quantum Hall transition; yet, its universal exponents still somewhat conflict between experiments, theoretical models, and numerical ansatzes. We study critical behaviors of quasi-2D Weyl semimetal systems with a finite thickness $L_z>1$, disorder, and external magnetic field $B_z$. By analyzing the scaling behaviors of the localization lengths and local density of states using recursive methods, we find that the finite thickness yields a deviation from the 2D quantum Hall universality ($L_z=1$ case) and a crossover toward the 3D Gaussian Unitary Ensemble  ($L_z\rightarrow \infty$ limit), potentially offering another cause of the discrepancy. Our work demonstrates the often-overlooked importance of auxiliary degrees of freedom, such as thickness, and that 3D quantum Hall physics is not merely a trivial finite-thickness extension of its 2D counterpart.
\end{abstract}

\maketitle

\section{Introduction} \label{sec:intro}

The quantum Hall effect characterizes the robust quantization of Hall responses in two-dimensional (2D) electron systems \cite{PhysRevLett.45.494, PhysRevLett.49.405, PhysRevB.23.5632, PhysRevB.25.2185, PhysRevB.27.6083}, resulting from the localization of electrons in disordered potentials \cite{PhysRev.109.1492, PhysRevLett.42.673, LEE198021, RevModPhys.80.1355}, as well as the discrete Landau levels. It is also one of the important platforms to experimentally verify the Anderson localization theory directly in an electron system. The quantum Hall transitions are quantum phase transitions between electronic phases with different TKNN integers and integer Hall responses\cite{li2009scaling, li2005scaling, PhysRevLett.61.1294, giesbers2009scaling, kaur2024universality, huckestein1995scaling, RevModPhys.80.1355, pruisken1984localization, pruisken1985dilute, arovas1997real, chalker1988percolation, zirnbauer1997toward, lee1993quantum, amado2011numerical, nuding2015localization, slevin2009critical, huckestein1990one, puschmann2019integer, zhu2019localization, ippoliti2018integer, gruzberg2017geometrically, klumper2019random, wang2017anderson, lee1996effects, huckestein1999integer, bhaseen2000towards, evers2001multifractality, evers2008multifractality, pook1991multifractality, yang1995interactions,zirnbauer2019integer}, where critical states at discrete energies or magnetic fields traversing the system are responsible for changes in the topological invariants and transport behaviors. Various theoretical studies, such as the percolation picture \cite{chalker1988percolation} or the renormalization group study \cite{pruisken1984localization, pruisken1985dilute, arovas1997real}, and numerical studies, including the Chalker-Coddington network models \cite{chalker1988percolation, zirnbauer1997toward, lee1993quantum, amado2011numerical, nuding2015localization, slevin2009critical}, continuous models \cite{zhu2019localization, ippoliti2018integer} and lattice models \cite{puschmann2019integer, zhu2019localization}, as well as experimental probes \cite{li2009scaling,giesbers2009scaling,li2005scaling,PhysRevLett.61.1294,kaur2024universality}, have been conducted on such critical phenomena. However, although generally consistent and aligned, subtle controversies and discrepancies remain over their critical exponents.

One of the most important critical behaviors in the quantum Hall transitions is the critical exponent $\nu$, which characterizes the divergence of the localization length at the critical point \cite{RevModPhys.80.1355}. The recent experiments measure $\nu=2.38\pm0.06$ on the hetero-structure \cite{li2009scaling}, $\nu=2.4\pm0.2$ on the multi-layer graphene \cite{kaur2024universality}, and $\nu=2.5\pm0.2$ \cite{giesbers2009scaling} on the single-layer graphene, respectively. On the other hand, although the early numerical results $\nu=2.3-2.4$ \cite{huckestein1995scaling, RevModPhys.80.1355} based on the Landau-space models \cite{huckestein1990one}, and the Chalker-Coddington network models \cite{lee1993quantum} coincide with the experimental results, newer numerical results with higher accuracy are significantly larger, such as $\nu=2.616\pm0.014$ and $\nu=2.566\pm0.013$ based on the Chalker-Coddington network models \cite{amado2011numerical, nuding2015localization}, $\nu=2.58\pm0.03$ based on the lattice models \cite{puschmann2019integer}, and $\nu=2.48\pm0.02$ based on the continuous model\cite{zhu2019localization}. To address such inconsistency, physicists proposed a geometrically disordered Chalker-Coddington network model \cite{gruzberg2017geometrically,klumper2019random}, which yields $\nu=2.37$ closer to the experimental results; the electron Coulomb interactions \cite{lee1996effects, huckestein1999integer}, as well as the spin-orbit coupling \cite{wang2017anderson}, have also been considered as alternative origins of such deviations.

Another remarkable aspect of the quantum Hall transitions is the multifractality of the critical wavefunctions \cite{ bhaseen2000towards, evers2001multifractality, evers2008multifractality, RevModPhys.80.1355,pook1991multifractality, yang1995interactions}. One way to quantitatively capture this feature is through the inverse participation ratios (IPRs) $P_q=\int d^d\vec{r} |\psi(\vec{r})|^{2q}$ \cite{wegner1980inverse, Castellani_1986, evers2000fluctuations}, where $\psi(\vec{r})$ is the wavefunction and $d$ is the dimension of the system. IPRs $P_q$ exhibit anomalous scaling behaviors with the system size $L$ at the critical point, i.e. $P_q \sim L^{-\tau_q}$. The exponents $\tau_q$ for 2D quantum Hall transitions have already been extensively investigated in theoretical studies \cite{bhaseen2000towards} and determined with high numerical precision based on various models \cite{evers2001multifractality, evers2008multifractality, pook1991multifractality, yang1995interactions}.

Quantum Hall physics was recently generalized to three dimensions (3D), model examples being topological semimetal slabs in perpendicular magnetic fields \cite{zhang2017evolution, zhang2019quantum, lin2019observation, nishihaya2018gate, uchida2017quantum, tang2019three, potter2014quantum, zhang2016quantum, li20203d, chang2021three}. Such an electron quantum system possesses pairs of Weyl points, discrete points in the Brillouin zone where the band gap closes. These Weyl points also serve as monopoles or anti-monopoles of the Berry curvature \cite{wan2011topological, armitage2018weyl}, dominate the low-energy behaviors such as transport \cite{nielsen1983adler, son2013chiral, xiong2015evidence}, and evolve into chiral modes in magnetic fields \cite{nielsen1983adler, son2013chiral}. The surfaces of topological semimetals also host open Fermi arcs \cite{xu2015discovery, lv2015observation, xu2016observation}, which, together with the bulk chiral modes, form closed cyclotron orbits that are essentially 3D in volume and support a fully quantized Hall effect in 3D, even in the presence of various types of disorder \cite{potter2014quantum, zhang2016quantum, li20203d, chang2021three}. The quantized Hall conductance plateaus and their in-between transitions have been experimentally established in topological materials, such as Cd$_3$As$_2$ \cite{zhang2017evolution, zhang2019quantum, lin2019observation, nishihaya2018gate, uchida2017quantum, tang2019three}.

Here, we study the critical behaviors in such 3D quantum Hall insulators in the presence of a perpendicular magnetic field $B_z$ and disorder $W$. Naturally, the 2D quantum Hall physics dominates when the model thickness $L_z$ is sub-leading to the localization length in the $\hat z$ direction; on the other hand, the physics gradually evolves toward 3D metal-insulator transitions in the large $L_z$ limit \cite{chalker1995three, wang1999metal, henneke1994anderson, su2017generic, slevin1997anderson, song2021delocalization, sbierski2015quantum}. Presumably, therefore, a crossover is in between, and we indeed observe a noticeable deviation away from the 2D quantum Hall transition universality as $L_z$ increases, consistent with semiclassical and percolation intuitions \cite{chalker1988percolation,song2021delocalization,isichenko1992percolation, sotta2003crossover, stauffer2018introduction}.

In particular, we study the in-plane scaling of the localization length for the critical exponent $\nu$ and the IPR for the fractal dimension $\tau_2$ at the critical points, as well as their dependence on the vertical thickness $L_z$. Indeed, as $L_z$ increases, we observe a departure from 2D universality and the advent of crossover to 3D physics, carrying various deviations that are not irrelevant even in the thermodynamic limit ($L_x, L_y \rightarrow \infty$ with finite $L_z$). Therefore, we emphasize the importance of $L_z$ in pinpointing the 2D quantum Hall transitions, which may help explain and reconcile the discrepancy between different models and studies.

We arrange the rest of the paper as follows: We present our theoretical models and numerical methods in Sec. \ref{sec:II}; In Sec. \ref{sec:2d}, we analyze the critical behaviors of 2D quantum Hall transitions ($L_z=1$) as a benchmark; we further move on to quasi-2D scenarios with finite thickness $L_z>1$, and summarize and discuss the emergent departure of critical behaviors in Sec. \ref{sec:3d}. Finally, we provide conclusive remarks and discussions on the potential implications of finite thickness $L_z$ and dimensionality on the discrepancies between various experimental and theoretical setups and results in Sec. \ref{sec:conclusion}.

\section{Models and methods}
\label{sec:II}

\subsection{3D quantum Hall physics in a Weyl semimetal slab}

We begin with the following tight-binding Hamiltonian on a 3D cubic lattice \cite{hosur2012friedel, zhang2016quantum}:
\begin{equation}
\begin{aligned}
H = &\sum_{x,y,z}\left[ \frac{it}{2} e^{-i\phi x}(c_{x,y-1,z+1}^{\dagger}c_{x,y,z} + c_{x,y-1,z-1}^{\dagger}c_{x,y,z}) \right. \\
&\left. + (-1)^zt_0c_{x,y,z+1}^{\dagger}c_{x,y,z}+h.c. \right] \\
&-(-1)^z\left[ \varepsilon_0 c_{x,y,z}^{\dagger}c_{x,y,z}- e^{\pm i \phi x}c_{x,y\pm1,z}^{\dagger}c_{x,y,z} \right. \\
&\left. - c_{x\pm1,y,z}^{\dagger}c_{x,y,z} \right]
+ w_{x,y,z} c_{x,y,z}^{\dagger}c_{x,y,z},
\end{aligned} \label{eq:ham}
\end{equation}
where the randomly and uniformly distributed $w_{x,y,z} \in [-W/2, W/2]$ stands for a quenched onsite disorder. $t$, $t_0$ are model parameters for inter-layer hoppings, and $\epsilon_0$ is a staggered onsite energy. Also, the $e^{i\phi x}$ phase factors represent the Landau gauge of a magnetic field $B_z$ of $\phi$ magnetic flux per lattice plaquette in the $\hat{z}$ direction. For simplicity, we set the unit convention as $\hbar=1$, $e=1$, and the lattice constant $a=1$ throughout this work, unless noted otherwise.

In the absence of the magnetic field $\phi$ and disorder $W$, we can simplify the Hamiltonian in the 3D momentum space:
\begin{equation}
\begin{aligned}
H(\vec{k})=&(\varepsilon_0-2\cos k_y-2\cos k_x)\sigma_z +(t_0-t\sin k_y)\sin k_z\sigma_y \\
&+\left [t_0(\cos k_z-1) - t \sin k_y (\cos k_z +1 )\right]\sigma_x, \\
\end{aligned}
\label{H_k}
\end{equation}
where $\sigma_{x,y,z}$ are the Pauli matrices denoting the sublattices (even and odd $z$), and $\vec{k}=(k_x, k_y, k_z)$. Consequently, a pair of Weyl points emerge at $(\pm k_0, 0, 0)$ in the Brillouin zone, where $k_0=\arccos(\varepsilon_0/2-1)$, with linear dispersion nearby:
\begin{equation}
\varepsilon (\vec{k})=\pm \sqrt{4\sin^2 k_0 \cdot (k_x-k_0)^2+4t^2 k_y^2+t^2_0 k_z^2},
\end{equation}
where the Fermi velocity $\vec{v}=(\sqrt{4\varepsilon_0-\varepsilon_0^2}, 2t, t_0)$ is adjustable through the corresponding parameters.

\begin{figure}
\centering
\includegraphics[width=1.0\linewidth]{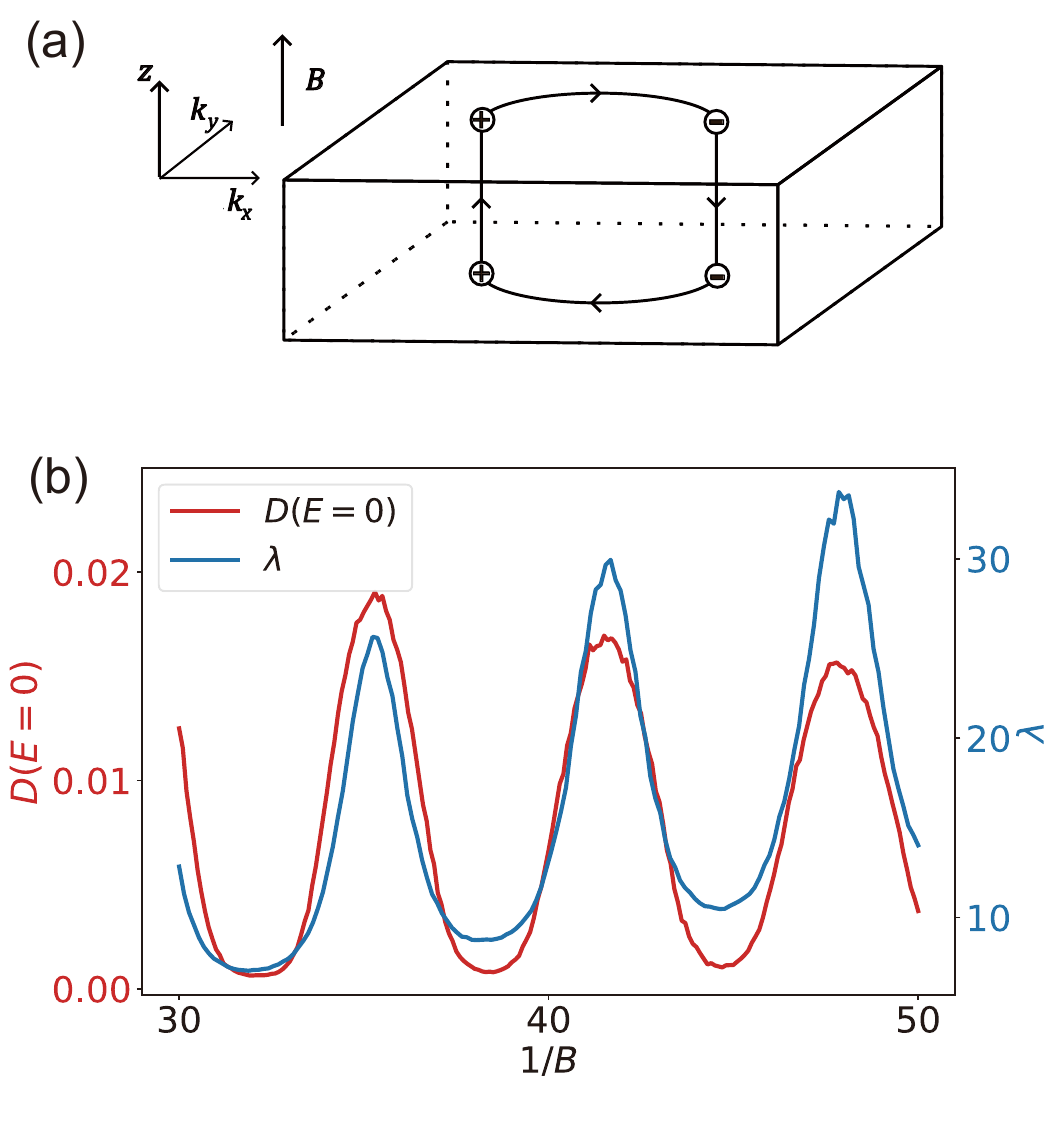}
\caption{(a) Illustration of the semiclassical electron orbit in a slab-shaped Weyl semimetal subject to a perpendicular magnetic field. The electron slides along the Fermi arc on the top surface, tunnels through the chiral Landau band to the opposite surface, and repeats this process on the bottom surface to form a closed cyclotron orbit. The circles denote the projections of the bulk Weyl points onto the surfaces, which also serve as the endpoints of the Fermi arcs; the $\pm$ symbols denote their respective chirality. (b) Both the averaged density of states $D(E=0)$ and the quasi-1D localization length $\lambda$ on disordered ($W=0.6$) systems of $L_x=20000$, $L_y=20$, and $L_z=11$ versus the inverse of the perpendicular magnetic field $1/B_z$ show clear signatures of quantum oscillations - a manifestation of the 3D quantum Hall effect \cite{zhang2016quantum}. }
\label{slab}
\end{figure}

Hereafter, we consider a slab-shaped system with open boundary condition (OBC) in the $\hat{z}$ direction with an odd $L_z$ and periodic boundary condition (PBC) in the $\hat y$ direction, unless specified otherwise. For the $\hat x$ direction, we implement either PBC or OBC with a large enough $L_x$. We set the model parameters as $t=1.0$, $t_0=0.6$, and $\varepsilon_0=3.7$. With such settings, the model supports the 3D quantum Hall effect, as reported in Refs. \onlinecite{zhang2017evolution, zhang2019quantum, lin2019observation, nishihaya2018gate, uchida2017quantum, tang2019three, potter2014quantum, zhang2016quantum, li20203d, chang2021three}. In the presence of the magnetic field $B_z$, the Weyl points quantize into a series of Landau bands:
\begin{equation}
\varepsilon_n^{\pm}(k_z)=
\begin{cases}
\pm\text{sgn}(n)\sqrt{2|n|v_xv_yl_B^{-2}+v_z^2k_z^2} & n=\pm1, \pm2, \dots\\
\pm v_z k_z & n=0,
\end{cases}
\end{equation}
where $l_B=\sqrt{\frac{\hbar}{eB}}$ is the magnetic length, and $\pm$ denotes the chirality of the Weyl points. Through the two $n=0$ Landau bands, gapless chiral modes, electrons can traverse between the top and bottom surfaces, and together with the surface Fermi arcs - the Weyl semimetal surface states, form closed cyclotron orbits in 3D \cite{potter2014quantum, zhang2016quantum}. Specifically, an electron can travel along the Fermi arc on one surface; once it reaches the endpoint of the Fermi arc, it tunnels to the opposite surface via the chiral Landau band in the bulk; it then continues its journey along the other Fermi arc, until it reaches the other Weyl point and tunnels back to the original surface through the chiral mode with opposite chirality; see Fig. \ref{slab}(a) for illustration. Despite plenty of spatial degrees of freedom along the thickness $\hat{z}$ direction, such a closed cyclotron orbit in 3D underlies the 3D quantum Hall effect, with full quantization as the 2D quantum Hall effect, as we show in Fig. \ref{slab}(b).

In the following, we consider properties of such Weyl semimetal slabs in the presence of a magnetic field and random disorder. Our motivation is two-fold: (1) theoretically, the 3D quantum Hall effect seems a naive finite $L_z$ generalization of its 2D counterpart, as one can recombine its Fermi arcs on the top and bottom surfaces into a closed 2D orbit; however, the large $L_z$ limit, where 3D physics entirely dominates, states otherwise; therefore, it is essential to study the universal behaviors of the 3D quantum Hall effect to debunk the former claim. (2) Unlike intrinsically 2D materials such as graphene, realistic 2D quantum Hall experiments, e.g., hetero-structures, commonly confine electrons to small yet finite thicknesses \cite{li2005scaling, li2009scaling, PhysRevLett.61.1294}, i.e., the penetration depth away from the interfaces; therefore, the impact of such $L_z$ dimensionality on quantum Hall universality may underlay practical values in explaining the observed discrepancies. Simple slab-geometry models, such as a quasi-2D cubic lattice in a magnetic field, yield Landau levels that are highly $ L_z$-dependent and usually too close to suppress inter-Landau level scattering. The 3D quantum Hall setup provides a viable alternative with sufficiently large and consistent gaps.

Alongside broadening the Landau levels, the disorder may introduce various scattering processes: (1) inter-valley scattering between the Weyl points, (2) inter-Landau-level scattering, and (3) random phase accumulation along the Weyl orbits \cite{zhang2016quantum}. For our purposes as stated above, we need to suppress the inter-valley scattering between the Weyl points and inter-Landau-level scatterings, which requires the model parameters to satisfy $\delta<W\lesssim\Delta$, where $\delta$ is the intrinsic width of Landau levels and $\Delta$ is the energy gap between the non-chiral Landau levels. Also, the magnetic field should not exceed an upper limit, above which the scattering between the Weyl nodes becomes non-negligible: $l_B \cdot \delta k\gg 1$, where $\delta k$ is the separation of the Weyl nodes. As we increase the ratio $L_z/l_z$, where $l_z$ is the dephasing length scale in the bulk (chiral modes), the system gradually undergoes a crossover from 2D to 3D. Due to the lack of translation symmetry, we resort to efficient numerical methods, which we discuss next.

\subsection{Recursive Greens function and transfer matrix method}

A 3D tight-binding model can be separated into $L_x$ slices of size $L_y \times L_z$; correspondingly, its Hamiltonian takes the following block tri-diagonal form:
\begin{equation}
H=\begin{pmatrix}
H_{1,1}&H_{1,2}&0&\ldots&0\\
H_{2,1}&H_{2,2}&H_{2,3}&\ldots&0\\
0&H_{3,2}&H_{3,3}&\ldots&0\\
\vdots&\vdots&\vdots&\ddots&\vdots\\
0&0&0&\ldots&H_{L_x,L_x}\\
\end{pmatrix},
\quad \Psi=\begin{pmatrix}
\Psi_1\\\Psi_2\\\Psi_3\\\vdots\\\Psi_{L_x} \\
\end{pmatrix},
\end{equation}
where $H_{ii}$ represents the terms within the $i^{th}$ slice, $H_{i+1,i}$  describes the hopping terms from the $i^{th}$ slice to the $(i+1)^{th}$ slice, and vice versa.

The transfer matrix method focuses on the decay rate $\lambda$ of a quasi-1D system ($L_x \gg L_{y,z}$), following the exponential decay of its wavefunctions \cite{mackinnon1983scaling}:
\begin{equation}
\Psi_n\sim\exp(-n/\lambda).
\end{equation}
To compute $\lambda$, we can re-express the eigenvalue equation as:
\begin{equation}
\begin{pmatrix}
\Psi_{n+1}\\\Psi_{n}\\
\end{pmatrix}
=M_n
\begin{pmatrix}
\Psi_{n}\\\Psi_{n-1}\\
\end{pmatrix},
\end{equation}
with the $n^{th}$ transfer matrix defined as:
\begin{equation}
M_n=\begin{pmatrix}
H_{n,n+1}^{-1}(E-H_{n,n})&-H_{n,n+1}^{-1}H_{n,n-1}\\
1 & 0\\
\end{pmatrix}.
\end{equation}
Therefore, the wavefunction propagation through the quasi-1D system is determined by the product of transfer matrices:
\begin{equation}
T_{L_x}=\prod_{n=1}^{L_x}M_n. \label{eq:ttmproduct}
\end{equation}

Define a Hermitian matrix $\Omega$:
\begin{equation}
\Omega=\lim_{L_x\rightarrow\infty}(T_{L_x}^\dagger T_{L_x})^{{1}/{2L_x}},
\end{equation}
we obtain the decay rate $\lambda$ via the smallest positive eigenvalue $\gamma_1$:
\begin{equation}
\lambda=1/\gamma_1,
\end{equation}
where $\pm\gamma_1,\pm \gamma_2, \dots, \pm \gamma_{L_y\times L_z}$ are the $L_y\times L_z$ pairs of eigenvalues of $\Omega$, $0<\gamma_1<\gamma_2<\dots<\gamma_{L_y\times L_z}$. We describe further details on the efficient computation of $\gamma_1$ in Appendix \ref{TM}.

For critical phenomena, it is customary for recursive methods to focus on large $L_x\rightarrow \infty$, which can be safely regarded as the thermodynamic limit and dropped from the expressions, and to analyze finite-size scaling with respect to $L_y$. We include results for different $L_x$ and thus the $L_y/L_x$ aspect ratio in the Appendix \ref{sec:appD}, which show that it does not affect the conclusion if it is large enough. In particular, for the critical exponent $\nu$ of the localization length $\xi$, we introduce the dimensionless quantity $\Gamma=L_y\gamma_1$ and data fit according to the finite size scaling \cite{mackinnon1981one,slevin1999corrections}:
\begin{equation}
\Gamma=f(u_0L_y^{1/{\nu}},u_1L_y^{-y}),
\end{equation}
where $f$ is a bivariate polynomial. $u_{0}$ and $u_{1}$ are polynomials of the tuning parameter $x$ that controls the distance away from the critical point $x_c$, e.g., the magnetic field strength or the Fermi energy. The critical exponent $\nu$ characterizes the divergence of the localization length $\xi$ at the critical point $x=x_c$:
\begin{equation}
\xi\sim|x-x_c|^{-\nu},
\end{equation}
and the other exponent $y>0$ contributes an irrelevant scaling parameter.

On the other hand, the recursive Green's function method \cite{mackinnon1985calculation,lewenkopf2013recursive} focuses on the Green's functions of the system, which also take a block matrix form:
\begin{equation}
G=\frac{1}{E+i\eta-H}=
\begin{pmatrix}
G_{1,1}&G_{1,2}&\ldots&G_{1,L_x}\\
G_{2,1}&G_{2,2}&\ldots&G_{2,L_x}\\
\vdots&\vdots&\ddots&\vdots\\
G_{L_x,1}&G_{L_x,2}&\ldots&G_{L_x,L_x}\\
\end{pmatrix},
\end{equation}
where $\eta\rightarrow 0^+$ is a small positive real number introduced to remove singularities and present a level broadening. The diagonal blocks of the Green's functions are:
\begin{eqnarray}
G_{n,n}&=&\left(E+i\eta-H_{n,n}-H_{n,n-1}G^L_{n-1,n-1}H_{n-1,n} \right.\nonumber \\
& & \left. -H_{n,n+1}G^R_{n+1,n+1}H_{n+1,n}\right)^{-1},
\end{eqnarray}
where $G^L$ and $G^R$ are the left and right Green's functions following the recursive relations:
\begin{equation}
\begin{aligned}
G^L_{n,n}=(E+i\eta-H_{n,n}-H_{n,n-1}G^L_{n-1,n-1}H_{n-1,n})^{-1},\\
G^R_{n,n}=(E+i\eta-H_{n,n}-H_{n,n+1}G^R_{n+1,n+1}H_{n+1,n})^{-1},\\
\end{aligned}
\end{equation}
for efficient calculations.

The local density of states (LDOS) $\rho(n, m, l)$, $n, m, l=1, 2, \dots, L_{x, y, z}$, are obtainable through the Green's functions:
\begin{equation}
\rho(n,m,l)=-\frac{1}{\pi L_xL_yL_z}\text{Im} G_{n,n}(m, l;m, l),
\end{equation}
where $G_{n,n}(m,l;m,l)$ are the diagonal matrix elements of $G_{n,n}$, corresponding to the site at $(n, m, l)$. Thus, beside the overall density of states $D(E)=\sum_{n, m, l}\rho(n, m, l)$, we can obtain the IPR:
\begin{equation}
P_2 = \sum |\psi(n, m, l)|^4 / \left(\sum |\psi(n, m, l)|^2\right)^2,
\end{equation}
defined for a single eigenstate through:
\begin{equation}
P_2=\left[ \sum \rho(n,m,l)^2 \right] / \left[\sum \rho(n,m,l)\right]^2,
\label{IPR_def}
\end{equation}
with a sufficiently small $\eta$, much smaller than the energy level spacings, to select out single eigenstates. The summations are over the spatial coordinates $n, m, l$. Such a recursive Green's function method is more efficient than exact diagonalization on large system sizes.

The scaling of the IPR offers a dimensional analysis of the states \cite{wegner1980inverse}:
\begin{equation}
P_2\sim L^{-\tau_2},
\end{equation}
where $\tau_2=d$ for a metallic system with extended wavefunctions and $\tau_2\rightarrow 0$ for an insulating system with localized wavefunctions. At the critical point, the wavefunction exhibits multifractality with $0<\tau_2<d$. Here, we vary the system sizes $L_x=L_y=L$ and fix $L_z$ to focus on the anomalous dimension in the $xy$ plane ($d=2$).

\section{Universality of 2D Quantum Hall transitions} \label{sec:2d}

For the special case of $L_z=1$, the Hamiltonian returns to a tight-binding model of free electrons on a 2D square lattice in a perpendicular magnetic field. Its dispersion quantizes into discrete Landau levels, which expand in width in the presence of disorder; nevertheless, most of the states remain localized, and only a few critical states are responsible for the quantized Hall conductance and quantum Hall transitions.

\begin{figure}
\centering
\includegraphics[width=1.0\linewidth]{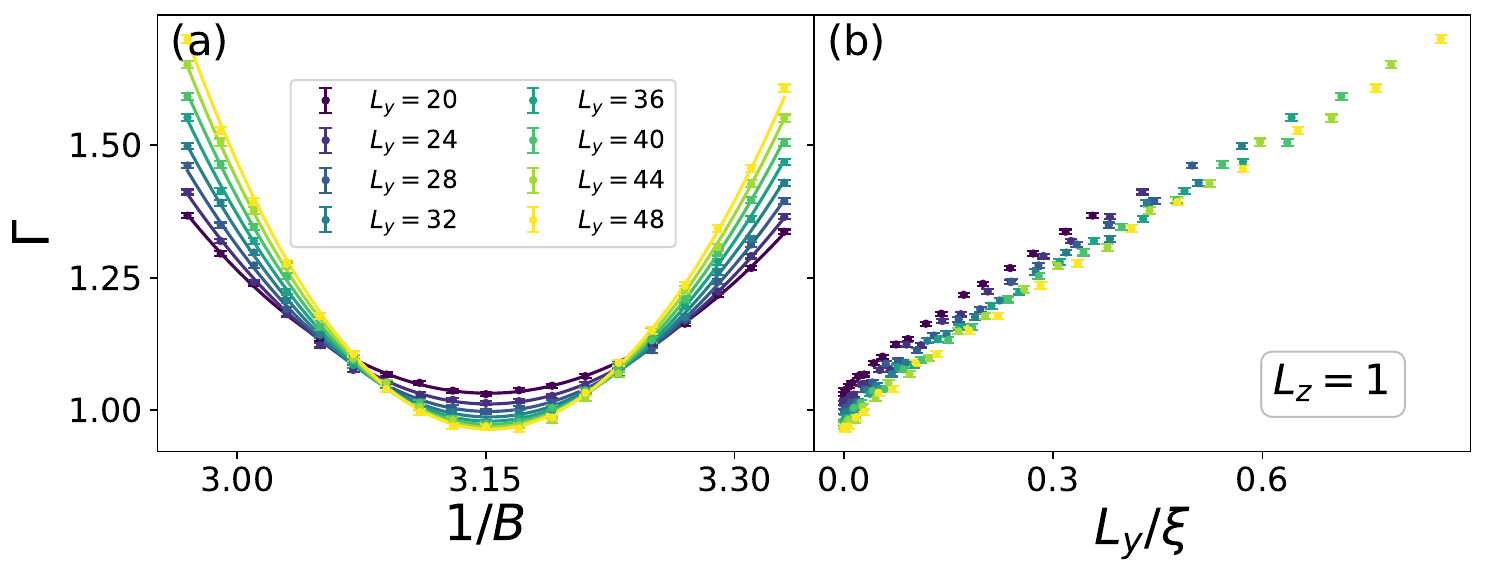}
\caption{The (inverse) dimensionless localization length $\Gamma$ for critical states in disordered 2D quantum Hall systems ($L_z=1$). (a) Results of $\Gamma$ around a critical magnetic field from the transfer matrix method (dots) display satisfactory fitting to the finite-size scaling (solid lines). (b) With the dimensionless quantity $L_y/\xi$, the data approximately collapses into a single curve; see further discussions in the main text and Appendix \ref{sec:appC}. Here, we estimate $\xi=|1/B-1/B_c|^{-\nu}$. We set the system width $L_y\in [20, 48]$, disorder strength $W=0.6$, and Fermi energy $E=0$ (at the original Weyl nodes).} \label{Lz=1_localization}
\end{figure}

\begin{figure}
\centering
\includegraphics[width=1.0\linewidth]{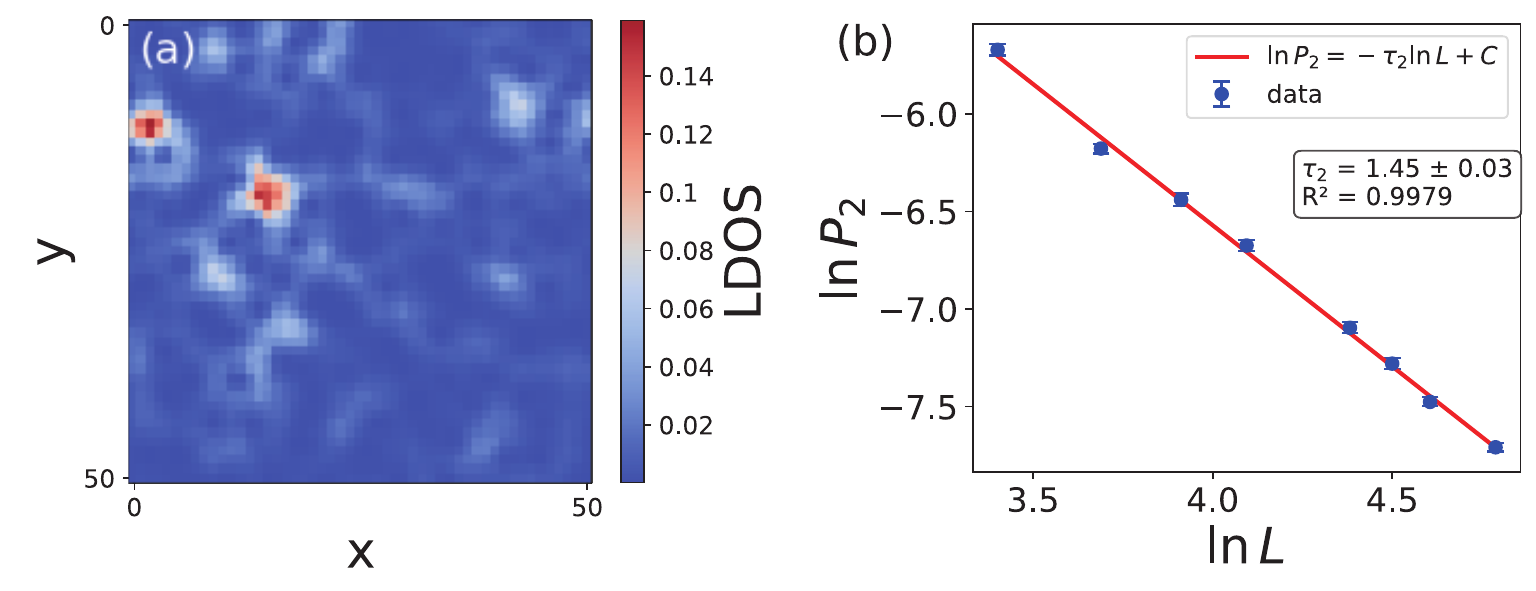}
\caption{The multifractality of wavefunctions at critical points in disordered 2D quantum Hall systems ($L_z=1$). (a) The LDOS of a single critical state, obtained from the recursive Green's function method, exhibits multifractal behaviors. Here, $\eta=0.01/L_xL_y$. (b) The scaling of the IPR $P_2$ with the system size $L\in [30,120]$ determines the fractal dimensions $\tau_2=1.45$ of the wavefunctions at the critical point. } \label{Lz=1_frac}
\end{figure}

To investigate the critical behaviors, we employ the transfer matrix method to calculate the (inverse) dimensionless localization length $\Gamma$ on sufficiently long systems $L_x=5\times 10^5$. After averaging over multiple disorder configurations, we summarize the numerical results of $\Gamma$ versus the (inverse) magnetic field ${1}/{B}$ and the dimensionless ratio ${L_y}/{\xi}$ in Fig. \ref{Lz=1_localization}.

In the vicinity of the critical point, which is further validated by the nice data collapse in Fig. \ref{Lz=1_localization}(b), $\Gamma$ scales as \cite{amado2011numerical, slevin1999corrections}:
\begin{equation}
\begin{aligned}
&\Gamma=\Gamma_c+\sum_{n=1}^{n_R}\alpha_{n}(u_0L_y^{\frac{1}{\nu}})^{n}+\sum_{m=1}^{n_I}\beta_m(u_1L_y^{-y})^m, \\
&u_0=\sum_{j=1}^{m_R}a_{j}(x-x_c)^{j}, \\
&u_1=\sum_{j=0}^{m_I}b_{j}(x-x_c)^{j}, \\
\end{aligned} \label{eq:fit}
\end{equation}
where we truncate the series at $n_R=4$, $n_I=1$, as well as the lowest orders $m_R=1$, $m_I=0$ for $u_{0,1}$ with $a_1=b_0=1$. Here, the tuning parameter $x$ denotes ${1}/{B}$. The resulting fit in Fig. \ref{Lz=1_localization}(a) is highly satisfactory with a reduced $\chi$ square of $\chi^2_\nu=\chi^2/N_d=1.06$, where $N_d$ is the degrees of freedom, and yields a critical exponent $\nu_{2D}=2.38\pm 0.03$ at the critical point $1/B_c=3.154$. Our results align approximately with previous studies, e.g., Refs.\onlinecite{huckestein1995scaling, puschmann2019integer} on relatively small system sizes, and provide a successful benchmark for our models and algorithms. Also, we find that odd-order terms like $\alpha_1(u_0L_y^{\frac{1}{\nu}})$ and $\alpha_3(u_0L_y^{\frac{1}{\nu}})^3$ are less dominant, thus the system behaviors near the transition point is symmetric with respect to the critical point, and the data approximately collapses into a single curve with the dimensionless quantity $L_y/\xi$ in Fig. \ref{Lz=1_localization}(b), consistent with the previous studies \cite{amado2011numerical, nuding2015localization, puschmann2019integer}. The variations away from a single-curve collapse across different $L_y$ are potentially due to the irrelevant terms in the scaling expression in Eq. \ref{eq:fit}, which we have retained in our data collapse. With the removal of such irrelevant contributions, or a weaker disorder, e.g., $W=0.1$ as in Appendix \ref{sec:appC}, the data can more consistently collapse into a single curve. On the other hand, our results are slightly lower in value than those of recent studies on larger system sizes, which we attribute to the finite-size effect; indeed, Refs. \onlinecite{zirnbauer2019integer} and \onlinecite{puschmann2019integer} argue that the critical exponent $\nu$ becomes larger when larger system sizes are employed. Nevertheless, instead of focusing on larger system sizes for more quantitative values of such critical exponents, we focus on the influence and tendency of quantum Hall universality due to a finite thickness $L_z$, and practical yet consistent system sizes in the $xy$ planes across our 2D and quasi-2D systems.

In addition to the localization length $\xi$ and $\Gamma$, we also calculate the LDOS of the disordered 2D quantum Hall system at the critical point using the recursive Green's function method with $\eta=0.001/L_xL_yL_z$, which is significantly smaller than the corresponding level spacings and guarantees the focus on single eigenstates. An example is in Fig. \ref{Lz=1_frac}(a), showing multifractal features. More quantitatively, we evaluate the IPR $P_2$ for various system sizes $L_x=L_y=L$, with PBC in the $y$ direction and OBC in the $x$ direction, where we include 10 extra layers on both ends and exclude them from our calculations in order to remove the potential impact of boundary effects \cite{obuse2008boundary} such as the potential chiral edge states and so on. Additionally, we calculate the average IPR and analyze its statistics across multiple disorder configurations. The scaling of the IPR, shown in Fig. \ref{Lz=1_frac}(b) with a highly satisfactory linearity $R^2=0.9979$, yields a characteristic critical dimension of $\tau_2=1.45\pm0.03$, well consistent with previous studies \cite{evers2008multifractality,pook1991multifractality}. Next, we expand our studies to quasi-2D scenarios with $L_z>1$.

\section{Universality in quasi-2D scenarios with finite thickness} \label{sec:3d}

\subsection{The 3D Gaussian Unitary Ensemble class}

Due to the magnetic field and thus the broken time-reversal symmetry, the Hamiltonian $H$ with random disorder falls into the Gaussian Unitary Ensemble (GUE) class \cite{dyson1962threefold, wegner1979mobility}. Unlike the GUE in 2D where most states are localized and only discrete critical states are responsible for non-local correlations and transport, the GUE in 3D can host finite regions of metallic phase \cite{chalker1995three, wang1999metal}, whose metal-insulator transitions to Anderson insulators, e.g., in strong disorder, and corresponding critical behaviors have been carefully studied, with a critical exponent $\nu_{3D}\sim 1.4$ for the localization length \cite{chalker1995three, wang1999metal, henneke1994anderson, su2017generic, slevin1997anderson, song2021delocalization, sbierski2015quantum}.

We note that we focus on quasi-2D systems with a slab geometry of thickness $L_z$, which means that when examining their critical behaviors, we scale only the $\hat{x}$ and $\hat{y}$ directions while keeping the thickness $L_z$ fixed. In contrast, $\nu_{3D}$ characterizes critical behaviors for scalings along all three spatial directions. Still, it points out the critical exponent $\nu$ in the $L_z\rightarrow \infty$ limit, where the $\hat{z}$ scaling becomes trivial. On the other hand, the critical exponent at $L_z=1$ returns to $\nu_{2D}$, the 2D quantum Hall universality. Therefore, we expect a crossover between $\nu_{2D}$ and $\nu_{3D}$ as $L_z$ increases.

Indeed, we may regard a quantum Hall system with a finite $L_z$ as effectively 2D if it is dominated by the average Berry curvature over $L_z$, which holds in the clean limit. However, the Berry curvature of a topological semimetal may depend drastically on their local potentials, which vary from $z$ to $z$ in the presence of disorder, making the distribution of Berry curvature across $z$ non-negligible. Such an argument holds best for slowly varying potentials (compared to wavefunctions), whose disruptive consequences to pure 2D physics, however, should carry over to local quenched disorder as well. Such a crossover is also apparent from a percolation theory perspective, an intuitive semiclassical description of quantum Hall domains and transitions in the presence of disorder \cite{chalker1988percolation, song2021delocalization}. The critical state of a 2D classical percolation system occurs at and only at a single point, i.e., the percolation probability $p=0.5$ \cite{isichenko1992percolation}. If the system possesses a finite thickness, however, the single critical point expands into a finite critical region $p_1<p<p_2$, which further broadens as the thickness increases \cite{isichenko1992percolation}, influencing its universal localization and distribution behaviors.

We note that previous studies have identified and analyzed related crossover behaviors of the GUE class governed by the inter-layer hopping parameter, i.e., hopping anisotropy, and the corresponding impact on the critical exponent $\nu_{3D}$ of the localization length in 3D \cite{chalker1995three, wang1999metal}, with scaling applied along all three spatial directions.

\subsection{Results of localization behaviors at $L_z>1$}

To investigate how the dimensionality influences the 2D quantum Hall universality, we adopt a quasi-2D slab geometry for our Hamiltonian $H$ in Eq. \ref{eq:ham}, with an odd thickness $L_z>1$ and OBC along the $\hat z$ direction, which ensures the Weyl semimetal physics in the clean limit, 3D quantum Hall physics in the magnetic field, as well as a straightforward comparison with the 2D limit ($L_z=1$). In the following, we study various values of $L_z$. In each trial, we keep $L_z$ fixed, analyze scalings along the $\hat{x}$ and $\hat{y}$ directions, and by summarizing the trend across $L_z$'s range, conclude its influences and control over the crossover. In both targets and settings, our settings are qualitatively distinctive from previous studies on the critical exponents $\nu_{3D}$ \cite{chalker1995three, wang1999metal}.

\begin{figure}
\centering
\includegraphics[width=1.0\linewidth]{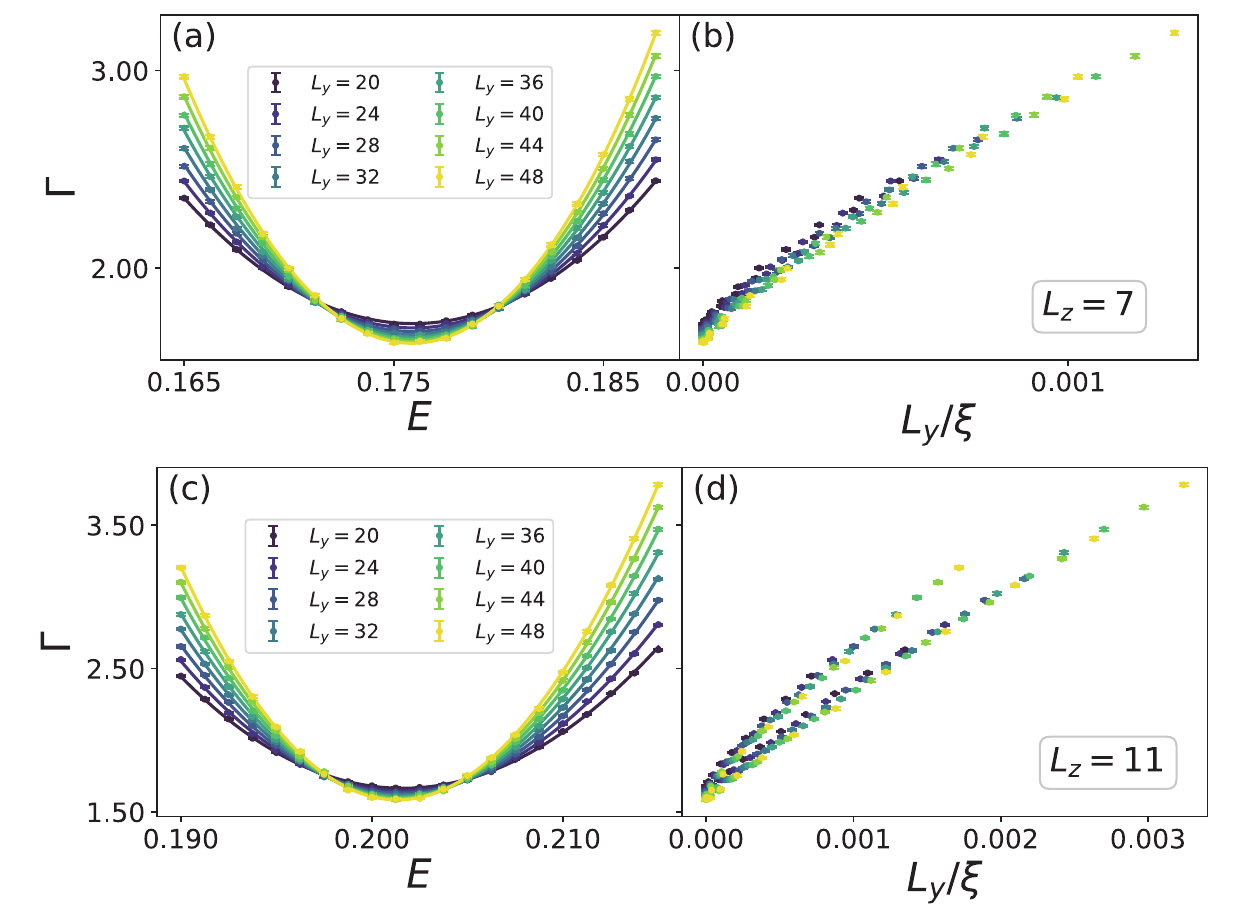}
\caption{The (inverse) dimensionless localization length $\Gamma$ versus the Fermi energy $E$ for quasi-2D systems in Eq. \ref{eq:ham} with magnetic field and disorder in slab geometry with thickness (a) $L_z=7$ and (c) $L_z=11$. $L_y\in [20, 48]$. The solid curves represent fits from the $L_y$ and $E-E_c$ scalings, as given by Eq. \ref{eq:fit}, with the resulting parameters listed in Table \ref{tab1}. Moreover, (b) and (d) are the corresponding data collapses near $E_c$, which increasingly reflect two separate branches, one for $E<E_c$ and one for $E>E_c$, as $L_z$ increases. }\label{Lz=15_localization2}
\end{figure}

First, we fix the magnetic field $1/B\sim3.2$, and vary $E$ for various $L_z>1$ values around their respective $E_c$. We employ the same numerical approach, the transfer matrix method, as in our benchmark $L_z=1$ systems in Sec. \ref{sec:2d}. After averaging over multiple disorder configurations on sufficiently long systems $L_x=5\times 10^5$, we obtain the (inverse) dimensionless localization length $\Gamma$ as shown in the left panels of Fig. \ref{Lz=15_localization2}. Subsequently, we fit the data using the scaling expression in Eq. \ref{eq:fit}, whose results are summarized in Table \ref{tab1}. Interestingly, as we have expected, the critical exponent $\nu$ witnesses a clear drop from $\nu_{2D}$ as $L_z$ increases, consistent with the crossover from 2D to 3D, $\nu_{3D}<\nu<\nu_{2D}$. In addition, unlike the 2D case in Sec. \ref{sec:2d}, the odd-order terms in Eq. \ref{eq:fit} become much non-negligible, reflecting the fact that the behaviors are no longer $x-x_c \rightarrow -(x-x_c)$ symmetric with respect to the critical point $x_c$. Indeed, the 3D metal-insulator transitions are actually characterized by two separate transitions at the mobility edges of a finite critical region, which are not related by any universal symmetry. As a result, the data do not collapse into a single curve as in Sec. \ref{sec:2d} and Appendix \ref{sec:appC}, but instead split into two branches, corresponding to the two sides of the critical point and irrespective of $L_y$ values, as we show in Fig. \ref{Lz=15_localization2}(b) and especially in Fig. \ref{Lz=15_localization2}(d). Thus, such an asymmetry in critical behaviors and bifurcation in data collapse may further imply, or even serve as an experimental signature, of a crossover from the 2D to 3D GUE class. We note the alternative explanation of contributions from multiple Landau levels across a sufficiently large range of $E$; however, we deem this an unlikely possibility as our $E$ range is much smaller than the corresponding Landau level spacing.

Note that we carry out the scaling studies as if there were a single critical point $E_c$, typically at the least localized point, instead of at the mobility edge of a finite metallic region as previously for $\nu_{3D}$ \cite{chalker1995three, wang1999metal, henneke1994anderson, su2017generic, slevin1997anderson, song2021delocalization, sbierski2015quantum}. This, together with the consistent 2D scaling, allows a straightforward comparison of the critical exponents as $L_z$ evolves from 1. Practically, it also plainly demonstrates the consequences of pure 2D narratives on systems that are essentially quasi-2D. Indeed, the observed deviation of the critical exponent from its 2D value offers an alternative explanation for discrepancies in certain experiments, which typically involve two-dimensional electron gas (2DEG) systems with finite thicknesses.

\begin{table}[htbp]
\centering
\begin{tabular}{c  c  c  c  c c c  c}
\hline
$L_z$ & $W$ & $\nu$ & $E_c$ &$\chi^2_{\nu}$\\
\hline
1 & 0.6 & $2.38\pm0.03$  & 0 &  1.06\\
7 & 0.6 & $2.37\pm0.02$  & $0.176$ & 0.82 \\
11 & 0.6 & $2.26\pm0.02$ & $0.201$ & 1.07\\
\hline
\end{tabular}
\caption{Model settings and $\nu$ fittings for variable $E$ at $1/B\sim 3.2$ and different $L_z$. }
\label{tab1}
\end{table}

Furthermore, as it is usually more customary to change the external magnetic field $B$ instead of the Fermi energy $E$ via gating, etc., we conduct another set of calculations with a fixed Fermi energy $E=0$ and varying magnetic field $B$. We summarize the results and discuss the potential ambiguity due to the magnetic-field-dependent penetration depth of surface orbits in Appendix \ref{sec:appB}.

\subsection{Results of distribution behaviors at $L_z>1$}

The critical behavior is also present in wavefunction distributions. We utilize the IPR, particularly its scaling behavior and exponent $\tau_2$, to quantify the spatial delocalization, localization, or multifractality of a target single wavefunction. For example, as a hallmark of 2D quantum Hall universality, previous numerical studies based on the Chalker-Coddington network model yield a fractal dimension $\tau_2\sim 1.43-1.48$ for 2D quantum Hall transitions \cite{evers2001multifractality, evers2008multifractality, pook1991multifractality}, consistent with our results on 2D lattice models with the recursive Green's function method at the end of Sec. \ref{sec:2d}.

Therefore, in addition to the critical behaviors of the localization length, we investigate the impact of dimensionality on the critical behaviors of the IPR at $E_c$ for quasi-2D systems with increasing thickness $L_z > 1$. Our model parameters, such as the fixed magnetic field $1/B\sim 3.2$ and disorder strengths $W$, are identical to those in Fig. \ref{Lz=15_localization2} and Table \ref{tab1}. Our model geometry is similar to that of IPRs at the end of Sec. \ref{sec:2d}, except that here the thickness $L_z$ becomes an odd integer greater than one. We emphasize again that, despite the current quasi-2D slab geometry, our scaling, $L_x=L_y=L$, is performed only along the $\hat{x}$ and $\hat{y}$ directions, which facilitates straightforward comparisons with the 2D quantum Hall transitions.

\begin{figure}
\centering
\includegraphics[width=1.0\linewidth]{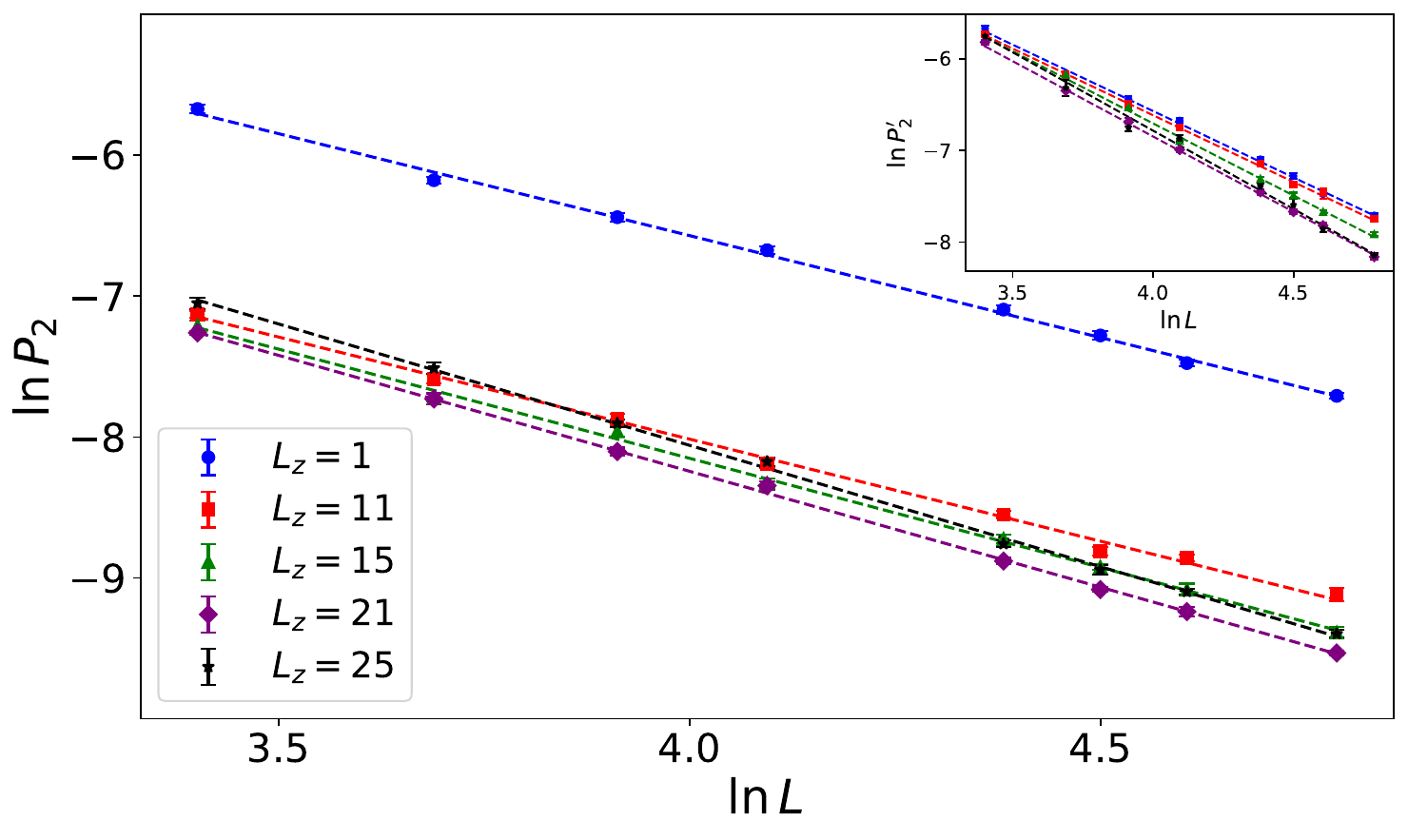}
\caption{The evolution of the IPR scaling exponent, i.e., the fractal dimension $\tau_2$ in the 2D $x-y$ plane, increases with the layer thickness $L_z>1$, indicating the advent of crossover from 2D to 3D universality in 3D quantum Hall systems. $L_x=L_y=L$, $L\in [30, 120]$. The main panel presents the IPR scaling of the entire system, while the inset shows the IPR scaling of the $z=(L_z+1)/2$ central layer. Clearly, all data displays excellent linearity (in the log-log plot), as indicated by the high $R^2$ value, as well as a change in slope as $L_z$ increases. Quantitatively, the resulting $\tau_2$ are listed in Table \ref{tab3}. }
\label{Fig_IPR}
\end{figure}

First, we calculate the IPR $P_2$ of the entire quasi-2D system and summarize our results in the main panel of Fig. \ref{Fig_IPR}. The 2D scaling of $P_2$ versus $L_x=L_y=L$ shows satisfactory power-law behaviors, with the exponent $\tau_2$, the fractal dimension, as summarized in Table \ref{tab3}. Clearly, an increase in $\tau_2$ with the thickness $L_z$ indicates a crossover from 2D to 3D, consistent with the further metallic state at $E_c$ for finite $L_z>1$, and complementing the localization-length behaviors above.

We note that the top ($z=L_z$) and bottom ($z=1$) surfaces of the model systems in slab geometry may make significant contributions to the IPRs, especially when the thickness $L_z$ is not sufficiently large and bulk contributions have not yet become dominant. To focus on the effects and marginal contributions from the bulk, we calculate the IPR $P'_2$ of a single layer located at $z=(L_z+1)/2$ for various $L_z$ thicknesses. We note that as $L_z \rightarrow \infty$, the physics of such central bulk layers becomes more dominant than that of the surfaces. The corresponding IPRs, shown in the inset of Fig. \ref{Fig_IPR}, yield the fractal dimensions presented in Table \ref{tab3}, which are consistent with those of the entire system and the emergent crossover from 2D to 3D. Finally, we note that previous studies have obtained a range of values for the anomalous dimension in 2D quantum Hall transitions using various scaling methods \cite{evers2001multifractality,evers2008multifractality,pook1991multifractality,yang1995interactions}. With consistent models and methods, we put aside such detailed differences and focus on the influence and tendency of $\tau_2$ due to the finite thickness $L_z$, as for the critical exponent $\nu$ in previous subsections.

\begin{table}[htbp]
\centering
\begin{tabular}{c  c c c  c c c}
\hline
\multirow{2}{*}{$L_z$} & \multirow{2}{*}{$W$} & \multirow{2}{*}{$E$} & \multicolumn{2}{c}{entire system} & \multicolumn{2}{c}{central layer}\\
 & & & $\tau_2$ & $R^2$ & $\tau_2$ & $R^2$\\
\hline
1  & 0.6 & 0 & $1.45\pm0.03$ & 0.9979 & $1.45\pm0.03$ & 0.9979\\
11 & 0.6 & 0.20 & $1.45\pm0.03$ & 0.9972 & $1.45\pm0.02$ & 0.9989\\
15 & 0.6 & 0.23 & $1.55\pm0.03$ & 0.9980 & $1.57\pm0.03$ & 0.9982\\
21 & 0.6 & 0.23 & $1.65\pm0.01$ & 0.9996 & $1.64\pm0.02$ & 0.9993\\
25 & 0.6 & 0.23 & $1.71\pm0.03$ & 0.9982 & $1.72\pm0.03$ & 0.9977\\
\hline
\end{tabular}
\caption{Model settings and $\tau_2$ fittings at critical $E$ and $1/B\sim 3.2$ for various $L_z$. }
\label{tab3}
\end{table}

\section{Conclusions and Discussions} \label{sec:conclusion}

To conclude, with the recursive Green's function and transfer matrix method, we study the critical phenomena, including the scaling behaviors of the localization lengths and the IPRs in quasi-2D Weyl semimetal systems with open boundary conditions and finite thickness along the $\hat z$ direction, disorder, and a perpendicular magnetic field. We discover that as $L_z$ increases, the critical exponents $\nu$ and $\tau_2$ shift away from the 2D quantum Hall values at $L_z=1$ and crossover towards those of the 3D magnetic metals. Simultaneously, an asymmetry develops between the two sides of the critical points $x_c$ in the data collapse of critical behaviors. As a side note, the 3D quantum Hall effect behind topological Weyl semimetals in a slab geometry, with its full Hall quantization and quantum oscillations \cite{zhang2017evolution, zhang2019quantum, lin2019observation, nishihaya2018gate, uchida2017quantum, tang2019three, potter2014quantum, zhang2016quantum, li20203d, chang2021three}, can sometimes be visualized as its 2D counterpart, i.e., cyclotron orbits in the $xy$ plane and extended quantum states across the thickness $\hat z$ direction. However, our results prove otherwise, and that the thickness may harbor further and richer physics beyond the 2D quantum Hall universality class \cite{NagaosaNanolett2022}.

Therefore, despite being often overlooked, the finite thicknesses, though seemingly insignificant, of 2DEG systems may leave a non-negligible imprint on their fundamental physics. Note that our conclusions regarding the thickness $L_z$ may generalize to other auxiliary degrees of freedom, such as sublattices and orbitals, as long as the disorder randomness can distinguish them, which is not a rare scenario in lattice models.

Experimentally, the 2DEG in a hetero-structure is realized through a quantum confinement $V(z)$ away from the target interface \cite{PhysRevLett.45.494, li2005scaling, li2009scaling, PhysRevLett.61.1294}, so that the localized electrons suffer a finite penetration depth into the 3D bulk. We now realize, however, that such a setup is naturally quasi-2D, and the thickness, i.e., on the order of the penetration depth, may impact the target critical behaviors, e.g., 2D quantum Hall physics. Further, such $z$ degrees of freedom become even more physically intertwined in disordered systems, as both the confinement and the disorder impact the electrons through the same local potentials $V(\vec{r})$. In contrast, electron systems that are fundamentally two-dimensional, such as graphene, are largely unaffected by such effects \cite{novoselov2004electric, novoselov2005two, giesbers2009scaling}. Also, multi-layer 2D materials provide a natural platform for crossover analysis and control through thickness variations \cite{taychatanapat2011quantum, novoselov2006unconventional, kaur2024universality}. Indeed, we note that, although potentially obscured by their uncertainties, the obtained critical exponents of various quantum systems \cite{li2009scaling, li2005scaling, PhysRevLett.61.1294, giesbers2009scaling, kaur2024universality}, as summarized in Sec. \ref{sec:intro}, align with the would-be shifts due to their thickness differences. Therefore, in addition to previously proposed explanations \cite{gruzberg2017geometrically, lee1996effects, huckestein1999integer, wang2017anderson, klumper2019random}, the quasi-2D dimensionality and thickness offer a distinctive, intuitive, and verifiable scenario for the discrepancy in 2D quantum Hall universality.

\section{Acknowledgments}

We acknowledge support from the National Key R\&D Program of China (Grant No.2021YFA1401900 \& No.2022YFA1403700), the National Natural Science Foundation of China (Grants No.12174008 \& No.92270102), and Shanghai Municipal Science and Technology Project (Grant No.25LZ2601100). \\

\section{Data availability}

The source codes and data that support the findings of this article are openly available \cite{github}.

\bibliography{ref.bib}

\appendix

\section{Algorithmic details of the transfer matrix method} \label{TM}

In this appendix, we discuss our method in practice for numerically obtaining the smallest positive eigenvalue $\gamma_1$ of the matrix $\Omega$, which is not readily and accurately computable directly from its definition. First, we define a series of intermediate matrices $M_k'$:
\begin{equation}
M_k'=M_{kN_0}M_{kN_0-1}\dots M_{(k-1)N_0+2}M_{(k-1)N_0+1}, k=1,2,\dots,N,
\end{equation}
where $N$, $N_0$ satisfy $NN_0= L_x$, and $N_0$ is a small integer. Then, we can rewrite the product of transfer matrices as:
\begin{equation}
T_{L_x}=\prod_{k=1}^{N}M'_k. \label{eq:A2}
\end{equation}

To overcome the difficulty of evaluating the product of transfer matrices, we can apply successive QR factorizations.\cite{mackinnon1983scaling} Specifically, we first apply a QR factorization to $M_1'$ as $M_1'=Q_1 R_1$, where $Q_1$ is a unitary matrix and $R_1$ is an upper-triangular matrix, and then recursively, $M_2'Q_1=Q_2R_2$, $M_3'Q_2=Q_3R_3$, etc., through their respective QR factorizations. Consequently, we can express the final product of transfer matrices in Eq. \ref{eq:ttmproduct} and Eq. \ref{eq:A2} as:
\begin{equation}
T_{L_x}=Q_N\prod_{k=1}^{N}R_k.
\end{equation}

To evaluate $\gamma_1$, we sum over (the logarithm of) the diagonal element $R_k(L_y\times L_z, L_y\times L_z)$ of the upper triangle matrix $R_k$ to obtain $c_n$:
\begin{equation}
c_n=\sum_{k=1}^{n} \ln R_k(L_y\times L_z,L_y\times L_z),
\end{equation}
and estimate $\gamma_1=c_N/NN_0$. In practice, we choose $N_0=7$.  We can estimate the relative uncertainty of the dimensionless quantity $\Gamma$ as $\sqrt{{M}/{L\Gamma}}$, following Ref. \onlinecite{zhang2005statistics}. Further, we average $\Gamma$ across multiple disorder configurations to reduce uncertainty. Consequently, the relative uncertainty of the data in Figs. \ref{Lz=1_localization} and \ref{Lz=15_localization2} is $\lesssim 0.5\%$.

Finally, once we obtain the localization lengths $\Gamma_i$ over various $x$ and $L_y$ settings, we fit the scaling expression $f_i(x, L_y)$ in Eq. \ref{eq:fit} according to the minimum $\chi$ square $\sum_{i}(\Gamma_i-f_i)^2/\sigma_i^2$ using the curve\_fit function from the Python Scipy library \footnote{The results also show little distinction from the fitting convention of $\sum_{i}(\Gamma_i-f_i)^2$. }. The corresponding uncertainties are obtained from the diagonal elements of the covariance matrix.

\section{Results and Discussions of varying $B$ measurements}\label{sec:appB}

\begin{figure}
\centering
\includegraphics[width=1.0\linewidth]{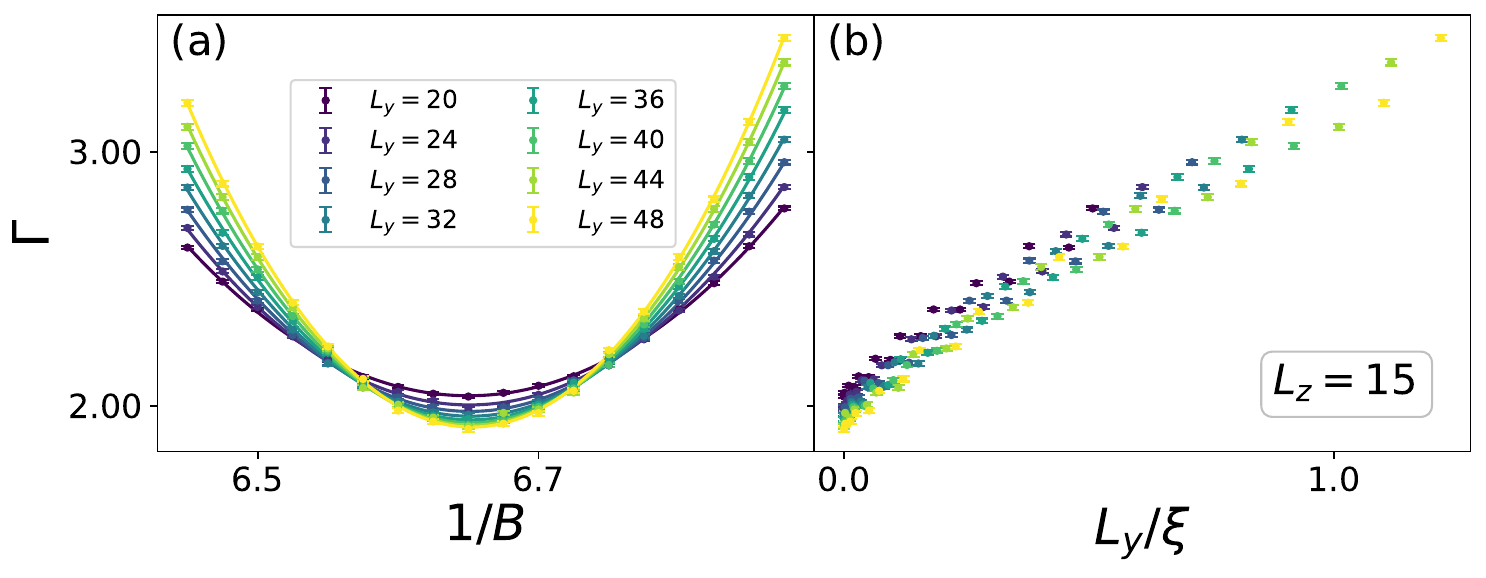}
\caption{(a) The localization length $\Gamma$ versus the inverse magnetic field ${1}/{B}$ for quasi-2D systems in Eq. \ref{eq:ham} in slab geometry with thickness $L_z=15$. $L_y\in [20, 48]$. The solid curves are fits from the $L_y$ and $1/B-1/B_c$ scalings, i.e., Eq. \ref{eq:fit}, with the resulting parameters listed in Table \ref{tab2}. (b) The corresponding data collapse near $1/B_c$ begins to fail as a single curve and shows two separate branches. }
\label{Lz=15_localization}
\end{figure}

In this appendix, we summarize the (inverse) dimensionless localization lengths $\Gamma$ and the corresponding scaling fits for thickness $L_z=15$, fixed $E=0$, and varying magnetic field $B$ in Fig. \ref{Lz=15_localization} and Table \ref{tab2}. The remaining model settings are identical to those for the fixed $B$ cases in the main text.

We note that semiclassically, the magnetic field controls the penetration depth of the surface orbits, and thus the effective bulk thickness below the bare $L_z$. The smaller the magnetic field $B_z$, the longer $l_B$ in the $xy$ plane, and the larger the size of the surface-bulk transitions, though it is hard to quantify precisely. This would probably matter less for large $L_z$; however, our $L_z$ is practically limited due to the computational cost. Also, the magnetic field alters $l_B$ and thus the dimensionless ratio $L_y/l_B$, whose largeness may further affect the derived critical exponent \cite{zirnbauer2019integer}. Therefore, the crossover tendency is more robust for a fixed $B$ in Fig. \ref{Lz=15_localization2} and Table \ref{tab1} in the main text than for a fixed $E$ (and descending $B$) in Fig. \ref{Lz=15_localization} and Table \ref{tab2}. Despite these influences, a tendency begins to emerge at a thickness of $L_z=15$ where the data collapse fails as a single curve and bifurcates into two branches. We leave more quantitative studies for larger system sizes and more carefully planned and consistently selected $B_c$ for future work.

\begin{table}
\centering
\begin{tabular}{c  c  c  c  c c c  c}
\hline
$L_z$ & $W$ & $\nu$ & $1/B_c$ & $\chi_\nu^2$ \\ \hline
1 & 0.6 & $2.38\pm0.03$ & $3.154$ & 1.06\\
1 & 0.1 & $2.36\pm0.02$ & $3.204$ & 0.91\\
15 & 0.6 & $2.40\pm0.03$ & $6.658$ & 1.02\\
\hline
\end{tabular}
\caption{Model settings and $\nu$ fittings for variable $B$ at $E=0$ and different $L_z$. }
\label{tab2}
\end{table}

\section{Results of 2D quantum Hall transitions with weaker disorder}\label{sec:appC}

\begin{figure}
\centering
\includegraphics[width=1.0\linewidth]{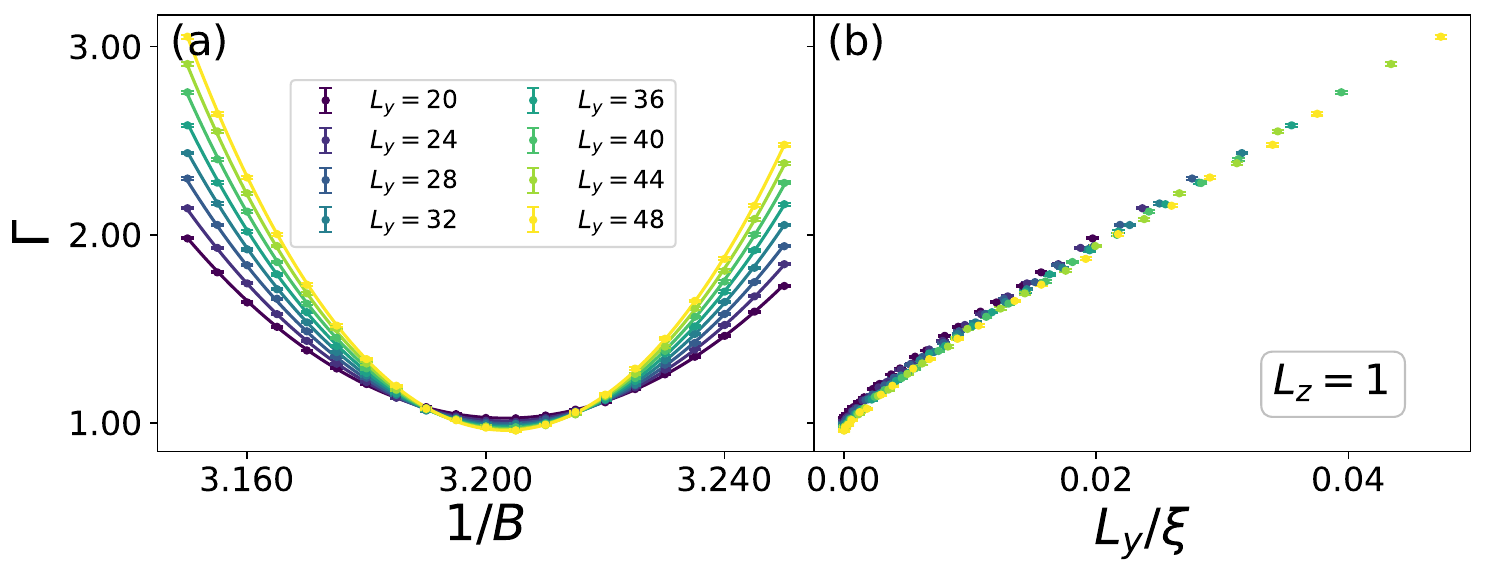}
\caption{The (inverse) dimensionless localization length $\Gamma$ for critical states in 2D quantum Hall systems ($L_z=1$) with weaker disorder $W=0.1$. (a) Results of $\Gamma$ around a critical magnetic field from the transfer matrix method (dots) display satisfactory fitting to the finite-size scaling (solid lines). (b) With the dimensionless quantity $L_y/\xi$, the data collapses into a single curve. We set the system width $L_y\in [20, 48]$ and Fermi energy $E=0$ (at the original Weyl nodes). } \label{Lz=1_localization_W=0.1}
\end{figure}

In this appendix, we summarize the results on the scaling behavior of the localization length and the IPR for the 2D system ($L_z=1$) with weaker disorder ($W=0.1$). The remaining model settings and methods are identical to those in Sec. \ref{sec:2d}. The fitting results of the calculated $\Gamma$ in Fig. \ref{Lz=1_localization_W=0.1}(a) yield a critical exponent $\nu_{2D}=2.36\pm0.02$ at the critical point $1/B_c=3.204$, summarized in Table \ref{tab2}. We also find that the data collapses, with better precision, into a single curve with the dimensionless quantity $L_y/\xi$ in Fig. \ref{Lz=1_localization_W=0.1}(b).

\begin{figure}
\centering
\includegraphics[width=1.0\linewidth]{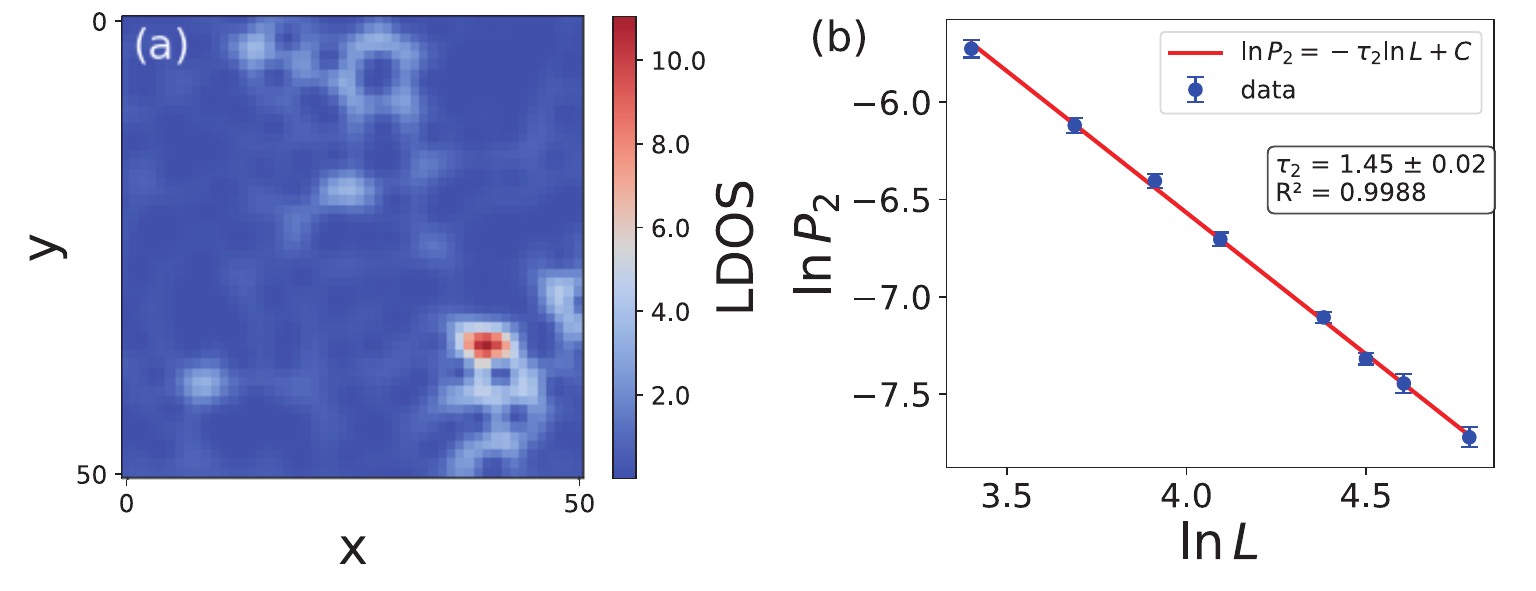}
\caption{The multifractality of wavefunctions at critical points in 2D quantum Hall systems ($L_z=1$) with weaker disorder $W=0.1$. (a) The LDOS of a single critical state, obtained from the recursive Green's function method, exhibits multifractal behaviors. Here, $\eta=0.01/L_xL_y$. (b) The scaling of the IPR $P_2$ with the system size $L\in [30,120]$ determines the fractal dimensions $\tau_2=1.45$ of the wavefunctions at the critical point. } \label{Lz=1_frac_W=0.1}
\end{figure}

Besides, the LDOS of the system at the critical point in Fig. \ref{Lz=1_frac_W=0.1}(a) shows the multifractal features, and the scaling of the IPR in Fig. \ref{Lz=1_frac_W=0.1}(b) yields a critical dimension $\tau_2=1.45\pm0.02$. Both results for the scaling behaviors of the localization length and the IPR agree well with those presented in Sec. \ref{sec:2d} of the main text, which correspond to stronger disorder.

\section{Results of different $L_y/L_x$ ratios}\label{sec:appD}

In this Appendix, we summarize in Tab. \ref{tab4} the (inverse) dimensionless localization lengths $\Gamma$ and the corresponding scaling fits for systems with a smaller $L_x=1\times10^5$. The remaining model settings are identical to those in the main text. The results for thickness $L_z=11$ are well consistent with those in the main text.

\begin{table}
\centering
\begin{tabular}{c  c  c  c  c c c  c}
\hline
$L_z$ & $W$ & $\nu$ & $E_c$ & $\chi_\nu^2$ \\ \hline
11 & 0.6 & $2.27\pm0.03$ & $0.201$ & 1.06\\
21 & 0.6 & $3.57\pm0.03$ & $0.229$ & 1.64\\
\hline
\end{tabular}
\caption{Model settings and $\nu$ fittings for variable $E$ at $1/B=3.2$ and different $L_z$. Here, we employ a smaller $L_x=1\times 10^5$ than the main text ($L_x=5\times 10^5$). }
\label{tab4}
\end{table}

On the other hand, the smaller $L_x$ allows us to study scenarios with a slightly larger $L_z$, which we also include in Tab. \ref{tab4}. Here, the finite-size scaling following Eq. 12 in the main text is relatively poor, as indicated by the $\chi_\nu^2$ value being significantly above 1. This dramatic shift indicates that the crossover is moving further into 3D, and the analysis based on a single critical point (2D physics) begins to show apparent deviations, given the actual expansion into a finite critical region (3D physics). Finally, we note that the $\tau_2$ studies are independent of such presumptions and solely rely on the physics at a single $x$ ($B$ or $E$), such as the results across extended values of $L_z$ in Tab. II in the main text.

\end{document}